\documentclass[a4paper]{jpconf}
\usepackage{graphicx}
\begin{document}
\title{Exploring QCD Matter at High Baryon Density}
\author{J. D. Brandenburg}

\address{
Shandong University, Rice University, Brookhaven National Laboratory
}

\ead{jbrandenburg@bnl.gov}

\begin{abstract}
This contribution presents a brief summary of the recent past efforts to experimentally explore the QCD phase diagram at high baryon chemical potentials through heavy-ion collisions. A few measurements are highlighted to present the current status in the search for a first-order phase transition, for a possible critical endpoint, and for evidence of chiral symmetry restoration. Finally, the outlook for the ongoing beam energy scan II program and future experiments at the FAIR complex are discussed. 
\end{abstract}

\section{Introduction}
The Facility for Anti-proton and Ion Research (FAIR) is a pioneering new accelerator facility that will provide access to the exotic types of matter present under extreme temperature and density like those that may be found in compact stars, stellar explosions, and in the early universe~\cite{collaboration_challenges_2016,senger_heavy-ion_2017}. 
In principle, one could learn everything about the various phases of nuclear matter from the theory of Quantum Chromodynamics (QCD). 
However, in practice this is not yet possible since direct QCD calculations are only viable for special cases. 
Specifically, lattice QCD calculations, which are applicable for large temperatures ($T$) and for low baryon chemical potentials ($\mu_B$), have indicated that hadronic matter transitions (through an analytic cross-over transition) into a deconfined phase of strongly interacting quarks and gluons above a critical temperature ($T_C \approx 150$ MeV$/c$)~\cite{aoki_order_2006,brown_existence_1990,borsanyi_transition_2011}. 
However, many other fundamental questions about QCD matter still remain that cannot currently be answered by direct theoretical calculations. 
Therefore, experimental exploration of strongly interacting matter (so-called QCD Matter) is necessary to address the structure of the QCD phase diagram at moderate temperatures and at high baryon chemical potentials (See Fig.~1 for a schematic of the QCD phase diagram and the many possible phases of QCD matter). 

\begin{figure}[h]
    \centering
    \begin{minipage}[b]{0.95\textwidth}
        \begin{center}
        \includegraphics[width=0.99\textwidth]{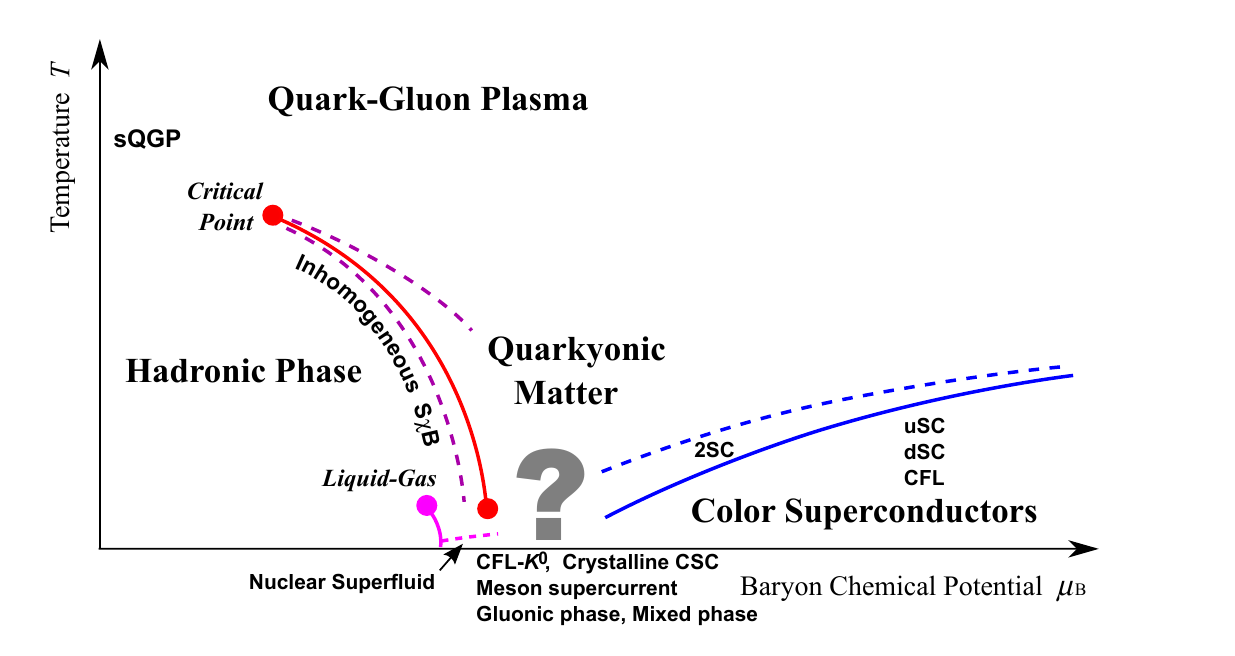}\hspace{2pc}%
        \caption{\label{label}A schematic of a possible QCD phase diagram showing several of the possible phases of QCD matter along with the phase boundaries separating them~\cite{fukushima_phase_2011}.  }
        \end{center}
    \end{minipage}
\end{figure}

One of the early success of the Relativistic Heavy Ion Collider (RHIC) was the creation of a dense, strongly interacting fluid of deconfined quarks and gluons, now called a quark-gluon plasma (QGP)~\cite{star_collaboration_experimental_2005,phenix_collaboration_formation_2005,al_phobos_2005,arsene_quarkgluon_2005}. 
Since that time considerable effort has been invested to characterize the QGP using high-energy heavy-ion collisions at RHIC and the Large Hadron Collider (LHC)~\cite{star_collaboration_azimuthal_2005,adare_jpsi_2011,adare_evolution_2012,cms_collaboration_jet_2012,cms_collaboration_observation_2012,the_atlas_collaboration_observation_2010,atlas_collaboration_measurement_2012,alice_collaboration_elliptic_2010}. 
At the nominal collision energies of the RHIC and the LHC ($\sqrt{s_{NN}}=200$ GeV and $\sqrt{s_{NN}}=2.76$ TeV), heavy-ion collisions produce a transient state of matter with $\mu_B\approx0$ that rapidly cools and expands, following a path through the QCD phase diagram similar to the path that the early Universe may have followed~\cite{boeckel_little_2012}.  
The impact parameter ($b$) and collision energy of heavy-ion collisions can be tuned to produce QCD matter that follows various trajectories through the phase diagram~\cite{cleymans_comparison_2006}. 
Higher values of $\mu_B$ can be accessed by colliding heavy-ions at lower collision energies (compared to nominal RHIC and LHC collision energies). 

This contribution briefly discusses and summarizes some of the important measurements from the recent past programs that have expanded our understanding of the QCD phase diagram and set the stage for future facilities like FAIR. 
The next sections discuss measurements related to 1) the search for a first-order phase transition between hadronic matter and a QGP at high $\mu_B$, 2) the search for a possible critical endpoint connecting the analytic cross-over transition at $\mu_B\approx 0$ to the first-order phase transition (if it exists) at higher $\mu_B$, and 3) the search for evidence of chiral symmetry restoration at sufficiently high $T$ and $\mu_B$. 
Lastly, the prospects for future measurements will be discussed with an emphasis on the ongoing RHIC beam energy scan (BES) II as well as the condensed baryonic matter (CBM) experiment and other FAIR programs.

\begin{figure}[h]
    \begin{minipage}[t]{0.38\textwidth}
        \begin{center}
        \includegraphics[width=0.99\textwidth]{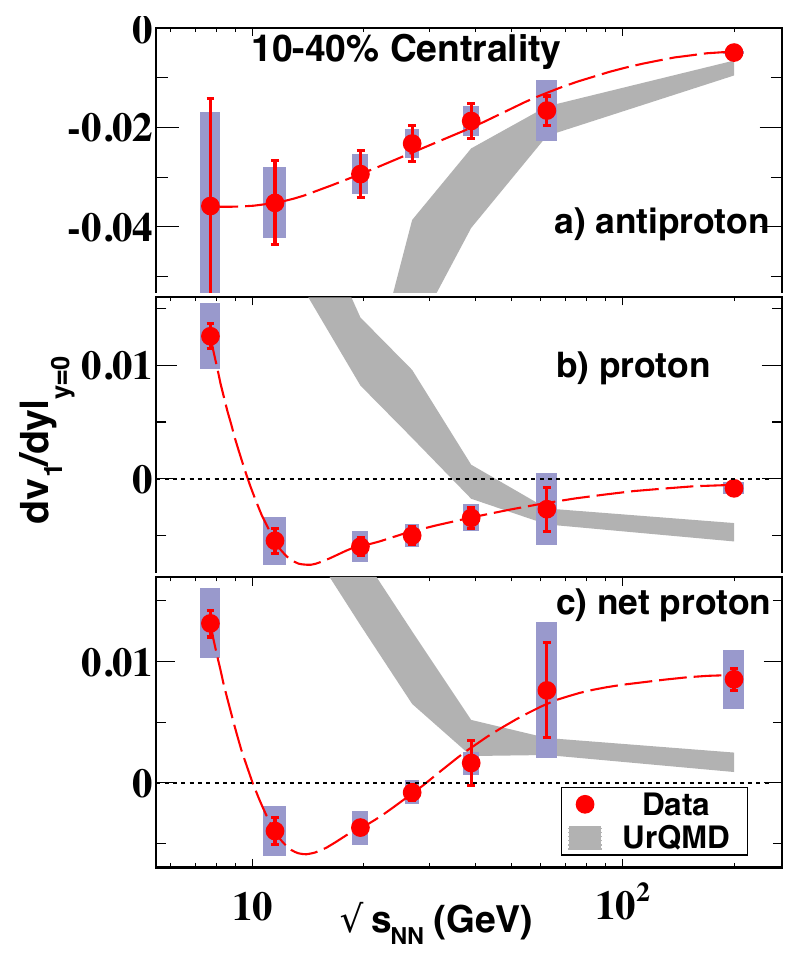}\hspace{2pc}%
        \caption{\label{label}The directed flow ($v_1$) at mid-rapidity ($y=0$) of anti-protons (top), protons (middle), and net-protons (bottom). The data are shown in red points with the UrQMD calculations shown in grey bands~\cite{star_collaboration_beam-energy_2014} }
        \end{center}
    \end{minipage}
    \hspace{0.07\textwidth}
    \begin{minipage}[t]{0.42\textwidth}
        \begin{center}
        \includegraphics[width=0.99\textwidth]{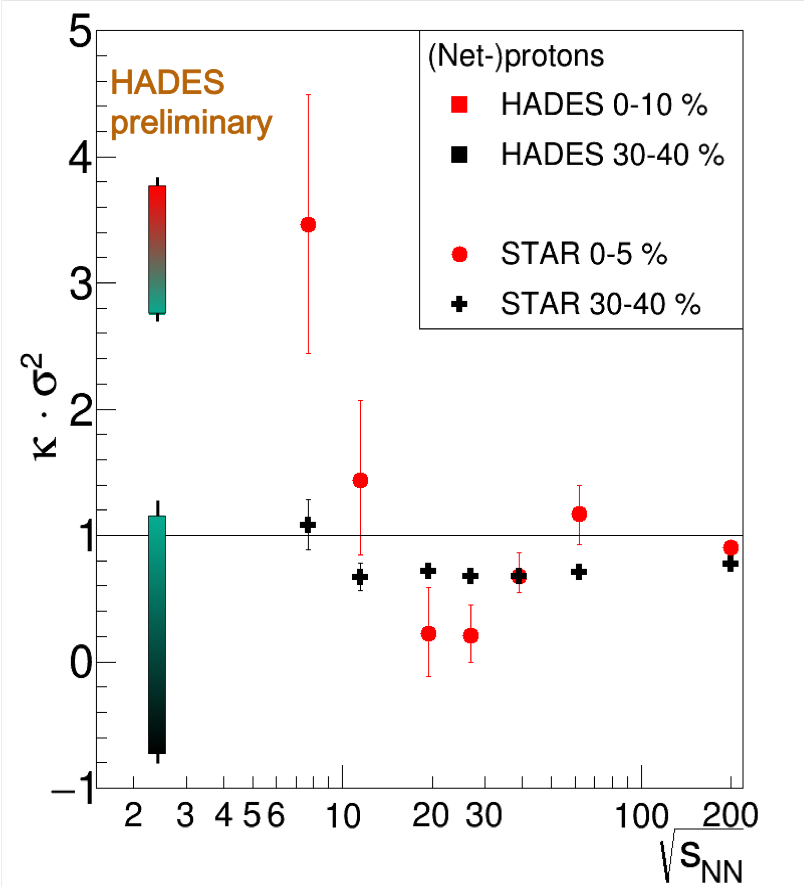}\hspace{2pc}%
        \caption{\label{label} The $\kappa \sigma^2$ for net protons measured from event-by-event proton and anti-proton multiplicities. The STAR and HADES data points are shown for two collision centralities~\cite{lorenz_overview_2017,luo_energy_2015}. The line at unity corresponds to the expected baseline for fluctuation according to Poisson statistics. }
        \end{center}
    \end{minipage}
\end{figure}

\section{ Search for a First Order Phase Transition and Possible Critical Endpoint }
Chiral effective models of QCD have suggested the possibility of a first order phase transition between hadronic matter and a deconfined QGP at finite $\mu_B$~\cite{fukushima_phase_2011}. Such a first-order phase transition is characterized by a spinodal region, i.e. a mechanically unstable coexistence region corresponding to a softest point in the equation of state. 
The directed flow of protons at mid-rapidity ($dv_1/dy|_{y=0}$), which is sensitive to the compressability of the system, is shown in Fig.~2 for several different collision energies from $\sqrt{s_{NN}}=$7.7 GeV to $\sqrt{s_{NN}}=$200 GeV~\cite{star_collaboration_beam-energy_2014}. 
A minimum in the $dv_1/dy|_{y=0}$ distribution for $p-\bar{p}$ (net-protons) is visible at a $\sqrt{s_{NN}}\approx10-20$GeV, which may be indicative of a softest point in the equation of state~\cite{steinheimer_directed_2014}, and therefore, may be evidence of a first-order phase transition. In addition to measurements of $v_1$, other notable measurements include quantum correlations (HBT), which can measure the effective size and shape of the emitting source. Pion HBT measurements show a change in the shape of the pion emitting source, from prolate at low energies to oblate at high energies, viewed from beside the beam, with the transition occurring near the same collision energy as the observed minimum in $v_1$ ~\cite{star_collaboration_beam-energy-dependent_2015,frodermann_evolution_2007,li_effects_2009}.

If a first order phase transition between hadronic and deconfined matter exists at finite $\mu_B$, then a critical endpoint should connect the smooth cross-over transition (at low $\mu_B$) with the first order phase transition~\cite{stephanov_signatures_1998}. A critical endpoint is characterized by a divergence in the correlation length of the system which is related to a divergence of susceptibilities. Ratios of susceptibilities can be accessed experimentally through event-by-event fluctuations in conserved quantities, such as the baryon number. Specifically, the product of the kurtosis and the sigma squared ($\kappa \sigma^2$) of event-by-event net proton multiplicities (net-protons are a proxy to the baryon number) has been proposed as an observable sensitive to the critical endpoint. Figure 3 shows the $\kappa \sigma^2$ for net-protons measured by STAR for two different centrality classes along with the same type of measurement from HADES~\cite{luo_energy_2015,lorenz_overview_2017}. The non-monotonic behavior visible in the STAR 0-5\% central data points (specifically the large rise in value near $\sqrt{s_{NN}}=7.7$ GeV) has been interpreted as a possible signature of the critical point~\cite{stephanov_signatures_1998}. However, such an interpretation generally requires that the measured $\kappa \sigma^2$ returns to unity (the Poisson baseline) at lower collision energies. Despite being at much lower $\sqrt{s_{NN}}$, the HADES measurement from 0-5\% central data has a value similar to the STAR data point at $\sqrt{s_{NN}}=7.7$ GeV. However, direct comparison of the two measurements is not trivial since the kinematic acceptance of the two experiments are quite different. 

\begin{figure}[h]
    \begin{minipage}[t]{0.40\textwidth}
        \begin{center}
        \includegraphics[width=0.99\textwidth]{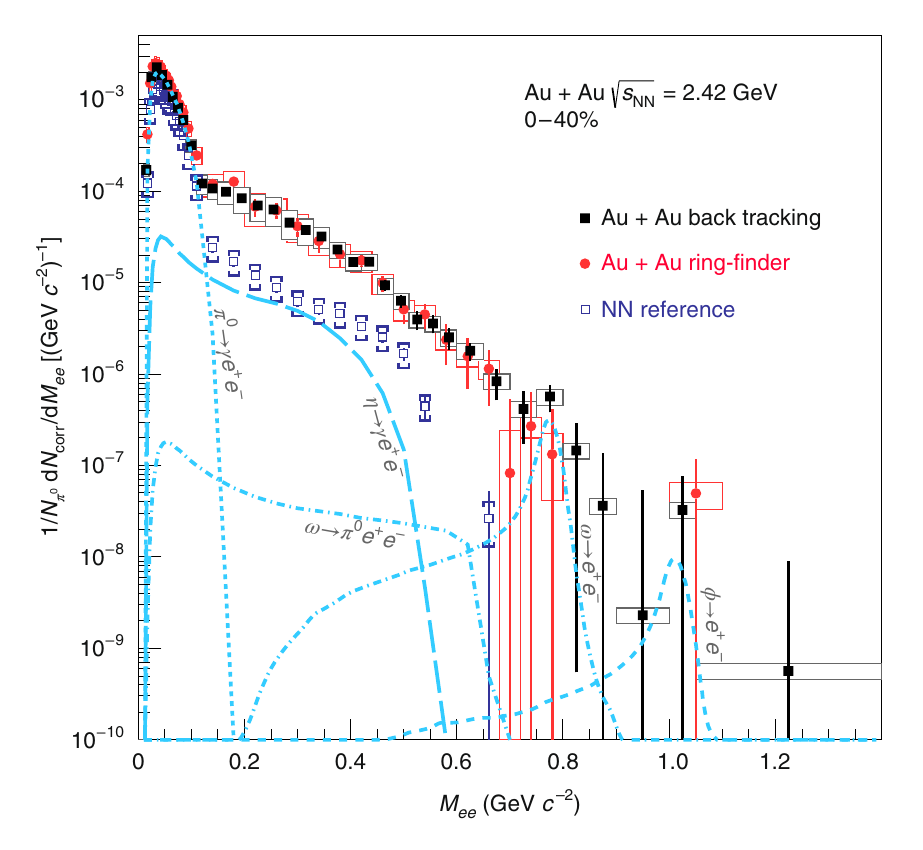}\hspace{2pc}%
        \caption{\label{label}The invariant-mass distributions of signal $e^+e^-$ pairs obtained from two different
analysis strategies (See ~\cite{hades_probing_2019}) measured by the HADES collaboration in Au+Au collisions at $\sqrt{s_{NN}}=$2.42 G$e$V. }
        \end{center}
    \end{minipage}
    \hspace{0.07\textwidth}
    \begin{minipage}[t]{0.50\textwidth}
        \begin{center}
        \includegraphics[width=0.99\textwidth]{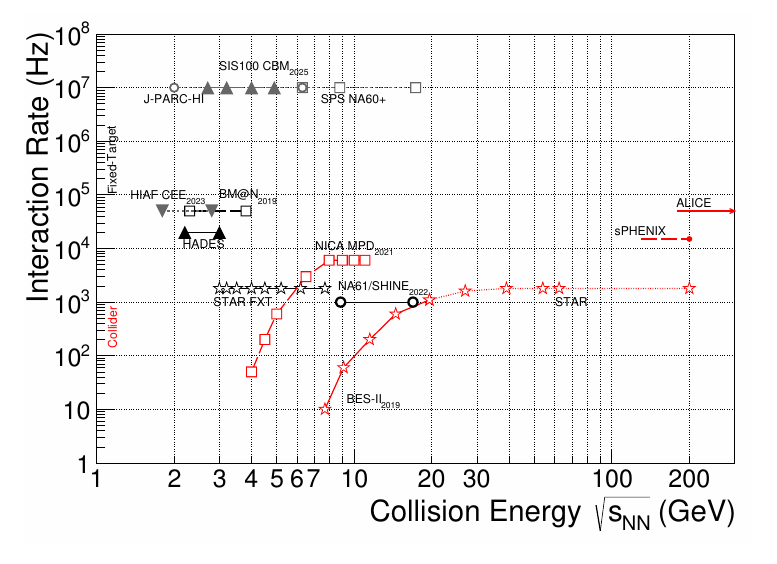}\hspace{2pc}%
        \caption{\label{label}The interaction rate versus center-of-mass collision energy of current and future facilities. Red points show collider mode experiments while black points show fixed-target experiments~\cite{galatyuk_future_2019}. }
        \end{center}
    \end{minipage}
\end{figure}

\section{Search for Chiral Symmetry restoration}
Theoretical predictions have indicated that the critical temperature for the transition from hadronic matter (with chiral symmetry broken) to a phase of matter with chiral symmetry restored may be roughly equal to the critical temperature for the deconfinement transition at low $\mu_B$~\cite{stephanov_signatures_1998}. At higher values of $\mu_B$ these two phase transitions may not coincide, making room for a phase of matter with chiral symmetry restored but still confined, a so-called Quarkyonic form of QCD matter. 
However, observing direct unambiguous evidence for chiral symmetry restoration is far from trivial. 
One proposed method is to observe the ``melting'' of the $\rho^0$ meson spectral distribution through measurements of lepton pairs~\cite{rapp_chiral_1999}.  Recently the HADES collaboration has conducted precision measurements of the dielectron spectrum in Au+Au collisions at $\sqrt{s_{NN}}=2.42$ GeV (see Fig.~4) which are useful for constraining the $\rho^0$ meson spectral function and for pristine measurement of the temperature of the produced fireball~\cite{hades_probing_2019,rapp_thermal_2016}. However, an unambiguous observation of chiral symmetry restoration via dileptons requires, in addition to the $\rho^0$ meson, measurement of its chiral partner, the $a_1$. Since the $a_1$ cannot be easily observed directly, the mixing between the vector and axial-vector may provide an alternative route for directly observing chiral symmetry restoration. Though, current experiments are not capable of reaching the statistical and experimental accuracy needed for such a measurement.

\section{Looking Forward}
While progress has been made, definitive quantitative answers about the structure of the QCD phase diagram at high $\mu_B$ have proven elusive. 
In many cases marginal statistics limits the accuracy of existing measurements and their corresponding ability to constrain the possible physics at play.
For this reason, collecting significantly higher statistics is an essential part of any next generation experiment that hopes to provide clarification about the open questions that still remain about the QCD phase diagram. 
The ongoing RHIC beam energy scan II program is precisely aimed at this goal, of re-measuring the range of energies covered in the BES I, but with higher statistics and with a significantly upgraded STAR detector. 
The BES II program will also feature a STAR fixed target program that extends the energy reach of the energy scan down to $\sqrt{s_{NN}} \approx$ 2.5 GeV, allowing a larger search region to be covered. 

Unlike the modest improvement in statistics of the BES II program compared to the BES I program, the fixed target experiments being planned for the FAIR complex (SIS100 CBM) are designed for an interaction rate that is orders of magnitude larger than the current programs (See Fig.~5)~\cite{galatyuk_future_2019}. 
Such a large increase in interaction rate will help bring several currently impractical measurements within reach, specifically the statistics hungry dilepton measurements. The FAIR facility and the CBM experiment represent an enormous opportunity for exploration of the QCD phase diagram through high statistics measurement of heavy-ion collisions. 
\section{Acknowledgements}
This work was funded in part by the U.S. DOE Office of Science under the contract number DE-SC0012704, the Brookhaven National Laboratory LDRD 18-037, and by Shandong University. 

\section*{References}
\bibliographystyle{iopart-num}
\bibliography{main}

\providecommand{\newblock}{}
\begin{thebibliography}{10}
\expandafter\ifx\csname url\endcsname\relax
  \def\url#1{{\tt #1}}\fi
\expandafter\ifx\csname urlprefix\endcsname\relax\def\urlprefix{URL }\fi
\providecommand{\eprint}[2][]{\url{#2}}

\bibitem{collaboration_challenges_2016}
{CBM Collaboration}, Ablyazimov T {\em et~al.\/} 2016  Arxiv:1607.01487

\bibitem{senger_heavy-ion_2017}
Senger P 2017 {\em J. Phys.: Conf. Ser.\/} {\bf 798} 012062

\bibitem{aoki_order_2006}
Aoki Y, Endrodi G, Fodor Z, Katz S~D and Szabo K~K 2006 {\em Nature\/} {\bf
  443} 675--678

\bibitem{brown_existence_1990}
Brown F~R, Butler F~P, Chen H, Christ N~H, Dong Z, Schaffer W, Unger L~I and
  Vaccarino A 1990 {\em Phys. Rev. Lett.\/} {\bf 65} 2491--2494

\bibitem{borsanyi_transition_2011}
Borsanyi S, Endrodi G, Fodor Z, Hoelbling C, Katz S~D, Krieg S, Ratti C and
  Szabo K~K 2011 {\em J. Phys.: Conf. Ser.\/} {\bf 316} 012020

\bibitem{fukushima_phase_2011}
Fukushima K and Hatsuda T 2011 {\em Rep. Prog. Phys.\/} {\bf 74} 014001

\bibitem{star_collaboration_experimental_2005}
{STAR Collaboration} and Adams J 2005 {\em Nuclear Physics A\/} {\bf 757}
  102--183

\bibitem{phenix_collaboration_formation_2005}
{PHENIX Collaboration}, Adcox K {\em et~al.\/} 2005 {\em Nuclear Physics A\/}
  {\bf 757} 184--283

\bibitem{al_phobos_2005}
{PHOBOS Collaboration}, Back B~B {\em et~al.\/} 2005 {\em Nuclear Physics A\/}
  {\bf 757} 28--101

\bibitem{arsene_quarkgluon_2005}
{BRAHMS Collaboration}, Arsene I {\em et~al.\/} 2005 {\em Nuclear Physics A\/}
  {\bf 757} 1--27

\bibitem{star_collaboration_azimuthal_2005}
{STAR Collaboration}, Adams J {\em et~al.\/} 2005 {\em Phys. Rev. C\/} {\bf 72}
  014904

\bibitem{adare_jpsi_2011}
{PHENIX Collaboration}, Adare A {\em et~al.\/} 2011 {\em Phys. Rev. C\/} {\bf
  84} 054912

\bibitem{adare_evolution_2012}
{PHENIX Collaboration}, Adare A {\em et~al.\/} 2012 {\em Phys. Rev. Lett.\/}
  {\bf 109} 152301

\bibitem{cms_collaboration_jet_2012}
{CMS Collaboration} 2012 {\em Physics Letters B\/} {\bf 712} 176--197

\bibitem{cms_collaboration_observation_2012}
{CMS Collaboration} 2012 {\em Phys. Rev. Lett.\/} {\bf 109} 222301

\bibitem{the_atlas_collaboration_observation_2010}
{ATLAS Collaboration} 2010 {\em Phys. Rev. Lett.\/} {\bf 105} 252303

\bibitem{atlas_collaboration_measurement_2012}
{ATLAS Collaboration} 2012 {\em Physics Letters B\/} {\bf 710} 363--382

\bibitem{alice_collaboration_elliptic_2010}
{ALICE Collaboration} 2010 {\em Phys. Rev. Lett.\/} {\bf 105} 252302

\bibitem{boeckel_little_2012}
Boeckel T and Schaffner-Bielich J 2012 {\em Phys. Rev. D\/} {\bf 85} 103506

\bibitem{cleymans_comparison_2006}
Cleymans J, Oeschler H, Redlich K and Wheaton S 2006 {\em Phys. Rev. C\/} {\bf
  73} 034905

\bibitem{star_collaboration_beam-energy_2014}
{STAR Collaboration}, Adamczyk L {\em et~al.\/} 2014 {\em Phys. Rev. Lett.\/}
  {\bf 112} 162301

\bibitem{lorenz_overview_2017}
Lorenz M 2017 {\em Nuclear Physics A\/} {\bf 967} 27--34

\bibitem{luo_energy_2015}
Luo X 2015 {\em Proceedings of 9th {International} {Workshop} on {Critical}
  {Point} and {Onset} of {Deconfinement} — {PoS}({CPOD}2014)\/}  019

\bibitem{steinheimer_directed_2014}
Steinheimer J, Auvinen J, Petersen H, Bleicher M and Stöcker H 2014

\bibitem{star_collaboration_beam-energy-dependent_2015}
{STAR Collaboration}, Adamczyk L {\em et~al.\/} 2015 {\em Phys. Rev. C\/} {\bf
  92} 014904

\bibitem{frodermann_evolution_2007}
Frodermann E, Chatterjee R and Heinz U 2007 {\em J. Phys. G: Nucl. Part.
  Phys.\/} {\bf 34} 2249--2254

\bibitem{li_effects_2009}
Li Q, Steinheimer J, Petersen H, Bleicher M and Stöcker H 2009 {\em Physics
  Letters B\/} {\bf 674} 111--116

\bibitem{stephanov_signatures_1998}
Stephanov M, Rajagopal K and Shuryak E 1998 {\em Phys. Rev. Lett.\/} {\bf 81}
  4816--4819

\bibitem{hades_probing_2019}
{HADES Collaboration} 2019 {\em Nat. Phys.\/}  1--6

\bibitem{galatyuk_future_2019}
Galatyuk T 2019 {\em Nuclear Physics A\/} {\bf 982} 163--169

\bibitem{rapp_chiral_1999}
Rapp R and Wambach J 2002 {\em Advances in {Nuclear} {Physics}\/}  1--205

\bibitem{rapp_thermal_2016}
Rapp R and van Hees H 2016 {\em Physics Letters B\/} {\bf 753} 586--590

\end{thebibliography}


\end{document}